\documentclass[a4paper,11pt]{article}

\usepackage[a4paper,left=2.73cm,right=2.7cm,top=3cm,bottom=3.5cm]{geometry}
\usepackage{amsmath,amssymb,graphicx,colortbl,xcolor,caption,subcaption,wrapfig}
\usepackage[colorlinks=true,linktocpage=true,linkcolor=blue,citecolor=blue]{hyperref}
\usepackage{cite}

\usepackage[all]{xy}
\definecolor{gray}{rgb}{.9,.9,.9}

\usepackage{caption}
\usepackage{subcaption}

\newcommand{\bea}{\begin{eqnarray}}
\newcommand{\beal}[1]{\begin{eqnarray}\label{#1}}
\newcommand{\eea}{\end{eqnarray}}
\newcommand{\be}{\begin{equation}}
\newcommand{\bel}[1]{\begin{equation}\label{#1}}
\newcommand{\ee}{\end{equation}}

\newcommand{\bit}{\begin{itemize}}
\newcommand{\eit}{\end{itemize}}
\newcommand{\ben}{\begin{enumerate}}
\newcommand{\een}{\end{enumerate}}

\newcommand{\sac}{\, , \qquad}

\newcommand{\thy}{t_\textrm{hyd}}
\newcommand{\teq}{t_\textrm{EoS}}
\newcommand{\Thy}{T_\textrm{hyd}}

\newcommand{\eqq}[1]{(\ref{#1})}
\newcommand{\fig}[1]{Fig.~\ref{#1}}

\newcommand{\bal}{\begin{align}}
\newcommand{\eal}{\end{align}}
\newcommand{\bse}{\begin{subequations}}
\newcommand{\ese}{\end{subequations}}

\def\phiM{\phi_M}
\def\O{{\mathcal{O}}}
\def\E{{\mathcal{E}}}
\def\P{{{P}}}

\def\Peq{{{P}}_{\rm eq}}
\def\Pt{{{P}}_T}
\def\Pl{{{P}}_L}
\def\V{{\mathcal{V}}}
\def\Veq{{\mathcal{V}}_{\rm eq}}

\definecolor{gray}{rgb}{.9,.9,.9}

\makeatletter \@addtoreset{equation}{section}
\makeatother

\begin{document}

\begin{titlepage}

\hfill{ICCUB-16-028}

\vspace{1cm}
\begin{center}

	    { \LARGE{\bf Holographic Collisions in Non-conformal Theories}}
		
\vskip 45pt

{\large \bf Maximilian Attems,$^{1}$ Jorge Casalderrey-Solana,$^{1,2}$  \\ [2mm]
David Mateos,$^{1, 3}$ 
Daniel Santos-Oliv\'an,$^{4}$  \\ [2mm]
Carlos F. Sopuerta,$^{4}$ Miquel Triana$^{1}$  and  Miguel Zilh\~ao$^{1}$}
		
		\vspace{25pt}

		{$^{1}$ Departament de F\'\i sica Qu\`antica i Astrof\'\i sica \&  Institut de Ci\`encies del Cosmos (ICC), Universitat de Barcelona, Mart\'{\i}  i Franqu\`es 1, 08028 Barcelona, Spain.}\\
		\vspace{15pt}
		{$^{2}$ Rudolf Peierls Centre for Theoretical Physics, University of Oxford, 
		\\ 1 Keble Road, Oxford OX1 3NP, United Kingdom}\\
		\vspace{15pt}
		{$^{3}$ Instituci\'o Catalana de Recerca i Estudis Avan\c cats (ICREA), \\
		Passeig Llu\'\i s Companys 23, 08010 Barcelona, Spain.}\\
		\vspace{15pt}
		{$^{4}$ Institut de Ci\`encies de l'Espai (CSIC-IEEC), Campus UAB,\\
Carrer de Can Magrans s/n, 08193 Cerdanyola del Vall\`es, Spain.}
		
\vskip 10pt
\end{center}

\vspace{10pt}
\abstract{\normalsize
We numerically simulate gravitational shock wave collisions in a holographic model dual to a non-conformal four-dimensional gauge theory. We find two novel effects associated to the non-zero bulk viscosity of the resulting plasma. First, the hydrodynamization time increases. Second, if the bulk viscosity is large enough then the plasma becomes well described by hydrodynamics before the energy density and the average pressure begin to obey the equilibrium equation of state. We discuss implications for the quark-gluon plasma created in heavy ion collision experiments. 

}

\end{titlepage}

\tableofcontents

\hrulefill
\vspace{15pt}

\section{Introduction} 
The gauge/string duality, also known as holography, has provided interesting insights into the far-from-equilibrium properties of hot, strongly-coupled, non-Abelian plasmas that are potentially relevant for the quark-gluon plasma (QGP) created in heavy ion collision experiments (see e.g.~\cite{CasalderreySolana:2011us} for a review). Most notably, holographic models have shown that ``hydrodynamization", the process by which the plasma comes to be well described by hydrodynamics, can occur before ``isotropization", the process by which all pressures become approximately equal to one another in the local rest frame. 

All far-from-equilibrium holographic studies of hydrodynamization to date (see e.g.~\cite{Chesler:2010bi,Casalderrey-Solana:2013aba,Casalderrey-Solana:2013sxa,Chesler:2015wra,Chesler:2015bba,Chesler:2016ceu}) have been performed in conformal field theories (CFTs).\footnote{Near-equilibrium studies of non-conformal plasmas include 
 \cite{Buchel:2015saa,Janik:2015waa,Buchel:2015ofa,Rougemont:2015wca,Gursoy:2015nza,Gursoy:2016tgf,Attems:2016ugt,Janik:2016btb}. 
 Refs.~\cite{Craps:2015upq,Ali-Akbari:2016sms,Gursoy:2016ggq} study far-from-equilibrium dynamics in non-conformal but homogeneous plasmas, hence no hydrodynamic modes are present. 
 Ref.~\cite{Buchel:2016cbj} studies bulk-viscosity-driven hydrodynamics in a cosmological context.} 
To make closer contact with the QGP, it is  important to understand non-conformal theories. One crucial difference between the two cases is that in non-conformal theories  the equation of state, namely the relation between the energy density and the average pressure, is not fixed by symmetry, and hence it needs not be obeyed out of equilibrium. The relaxation process therefore involves an additional channel,  namely the evolution of the energy density and the average pressure towards asymptotic values related by the equation of state. We will refer to this process as ``EoSization" and once it has taken place we will say that the system has ``EoSized''. One purpose of this paper is to show that hydrodynamization can  occur before EoSization.

We will consider gravitational shock wave collisions in a five-dimensional bottom-up model \cite{Attems:2016ugt} consisting of gravity coupled to a scalar field with a non-trivial potential. At zero temperature, the dual four-dimensional gauge theory exhibits a Renormalization Group (RG) flow from an ultraviolet (UV) fixed point to an infrared (IR) fixed point. The source $\Lambda$ for the relevant operator that triggers the flows is responsible for the breaking of conformal invariance. The dual gravity solution describes a  domain-wall geometry that interpolates between two AdS spaces. We emphasize that our choice of model is not guided by the desire to mimic  detailed properties of Quantum Chromodynamics (QCD) but by simplicity: the UV fixed point guarantees that holography is on its firmest footing, since the bulk metric is asymptotically AdS; the IR fixed point guarantees that the solutions are regular in the interior; and turning on a source for a relevant operator is the simplest way to break conformal invariance.

In this paper we  focus on a concise comparison between  hydrodynamization and  EoSization. Further details will be given in \cite{more}.

\section{The model}
The action for our Einstein-plus-scalar models is 
\be
\label{eq:action}
S=\frac{2}{\kappa_5^2} \int d^5 x \sqrt{-g} \left[ \frac{1}{4} \mathcal{R}  - \frac{1}{2} \left( \nabla \phi \right) ^2 - V(\phi) \right ] \,.
\ee
The potential 
\be
\label{eq:pot}
L^2 V=-3 -\frac{3}{2} \phi^2 - \frac{1}{3} \phi^4 + \left( \frac{1}{3 \phi_M^2} +  \frac{1}{2 \phi_M^4}\right) \phi^6-\frac{1}{12 \phi_M^4} \phi^8 
\ee
depends on one parameter $\phi_M>0$ in such a way that it possesses a maximum at $\phi=0$ (the UV) and a minimum at \mbox{$\phi=\phi_M$} (the IR). $L$ is the radius of the corresponding AdS solution at the UV, whereas the radius of the IR AdS is 
\be
\label{eq:LIR}
L_{\rm IR} = \sqrt{- \frac{3}{V\left(\phi_M\right)}} = \frac{1}{1+ \frac{1}{6} \phi_M^2} L \, .
\ee
The decrease of the number of degrees of freedom along the flow is reflected in the fact that $L_{\rm IR} < L$.
The scalar field is dual to a scalar operator in the gauge theory, $\O$, with dimension $\Delta=3$ at the UV fixed point.
The full flow describing the vacuum of the theory is given  by 
\bse
\label{all}
\bal
\label{eq:dsvac}
ds^2 &= \frac{du^2}{u^2} + e^{2 A(u)} \left(-d z_+ d z_- + d{\bf x}_\perp^2\right) \,, \\
\label{eq:metricsol}
e^{2 A}&= \frac{\Lambda^2 }{\phi^2} \,
  \left(1- \frac{\phi ^2}{\phi _M^2} \right)^{\frac{\phi_M^2}{6}+1} \, 
  e^{-\frac{\phi ^2}{6}}  \,, \\
\label{eq:phisol}
\phi&= \frac{\Lambda \, u}{\sqrt{1+ \frac{\phi_0^2}{\phi_M^2} u^2}}  \,,\end{align}
\ese
where $z_\pm = t \pm z$ and $\phi_0=\Lambda L$. The Ward identity for the trace of the stress tensor reads
\be
\label{eq:TTrace0}
\left<T^{\mu}_\mu\right>= - \Lambda \left< \O \right> \,,
\ee
and we adopt a renormalization scheme such that \mbox{$\left<T_{\mu\nu}\right>= \left<\O \right>=0$} in the vacuum.  Henceforth we will omit the expectation value signs and work with the rescaled quantities 
\be
({\cal E}, \Pl, \Pt, \mathcal V )=
\tfrac{\kappa_5^2}{2 L^3} (-T^t_t, T^z_z, T^{x_\perp}_{x_\perp}, \O ) \,.
\ee
In these variables the Ward identity takes the form
\be
\E - 3 \bar \P= \Lambda \V \,,
\ee
where 
\be
\bar \P = \tfrac{1}{3} \left( \Pl + 2 \Pt \right) 
\ee
is the average pressure. Out of equilibrium the average pressure is not determined by the energy density because the scalar expectation value $\V$  fluctuates independently. In contrast, in equilibrium $\V$    is determined by the energy density and the Ward identity becomes the equation of state
\be
\bar P = \Peq (\E) \,,
\ee
with
\be
\Peq (\E) = \tfrac{1}{3} \left[ \E -  \Lambda \Veq (\E) \right]\,.
\ee

\begin{figure}[t!!!]
\begin{center}
\includegraphics[width=.55\textwidth]{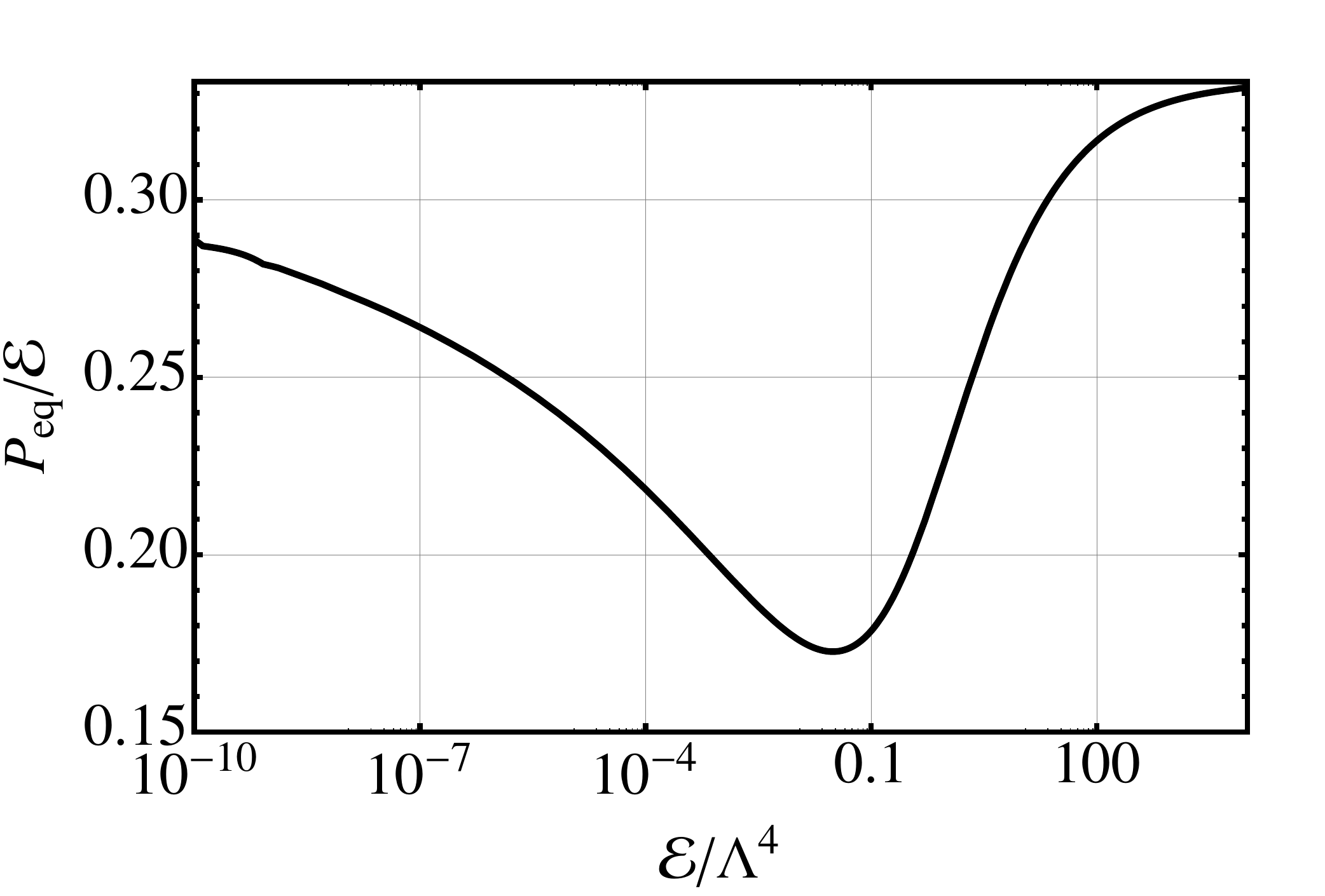}
\caption{\label{eos} Equilibrium pressure as a function of energy density for $\phiM=10$.}
\end{center}
\end{figure}

\begin{figure}[t!!!]
\begin{center}
\includegraphics[width=.55\textwidth]{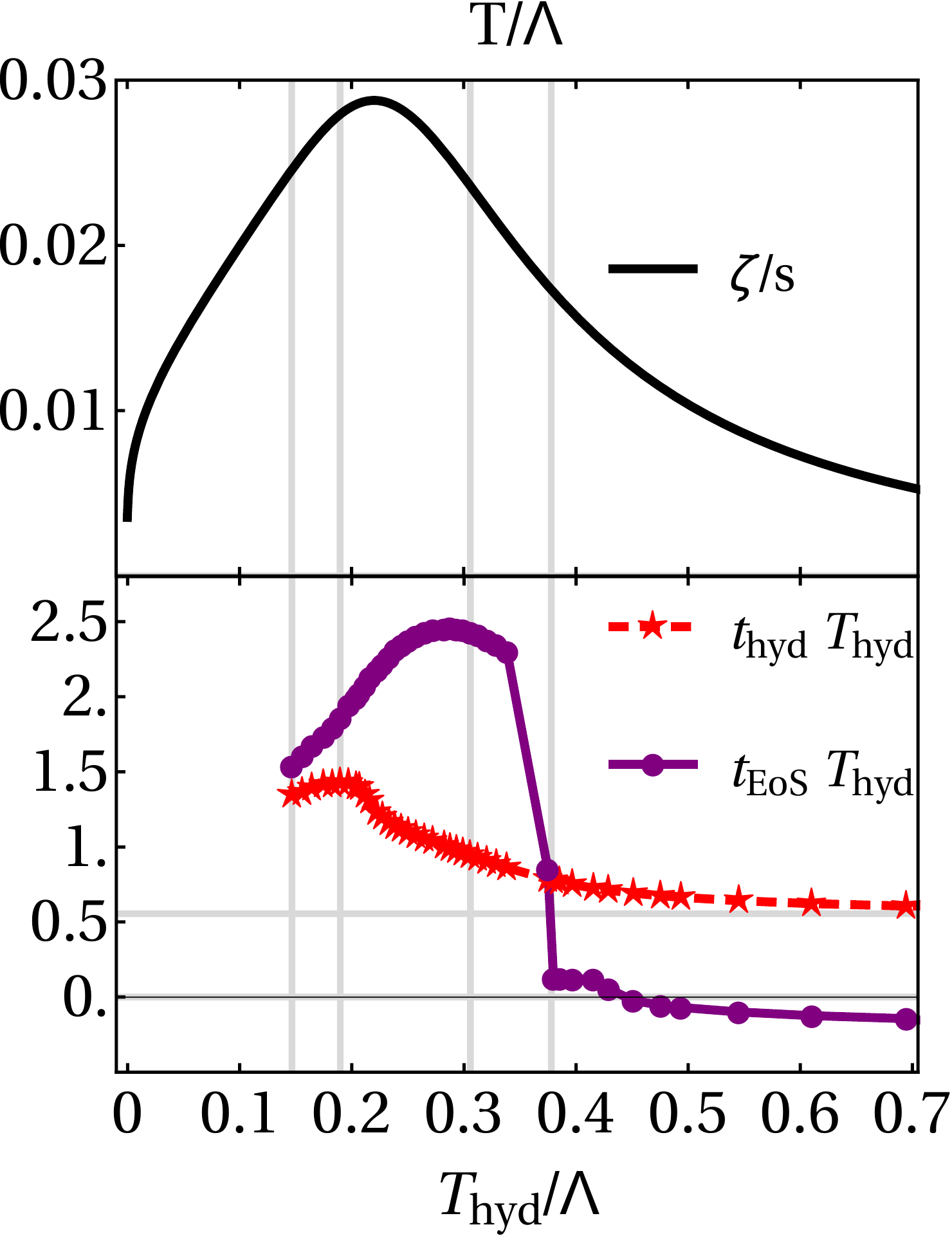}
\caption{\label{zeta} (Top panel) Bulk viscosity over entropy density as a function of temperature. (Bottom panel) Hydrodynamization and EoSization times as a function of the hydrodynamization temperature for collisions of $1/2\,$-shocks. The vertical grid lines lie at $T/\Lambda=\{ 0.15, 0.19, 0.31, 0.38 \}$ and mark, respectively,  the lowest value of $ \Thy/\Lambda$ that we have simulated, the maximum of $\thy \Thy$, the point with the largest ratio of $\teq/\thy$, and the intersection between the two curves. The bulk viscosity at these temperatures is $\zeta/s=\{0.025, 0.028, 0.023, 0.017 \}$. The top horizontal line indicates the result in a CFT, $\thy \Thy=0.56$.$^{\ref{note}}$}
\end{center}
\end{figure}

In this paper we will focus on $\phi_M=10$ (the dependence of the physics on this parameter will be presented in \cite{more}). The equilibrium pressure for this case is shown in \fig{eos}. As expected, both at high and low energies the physics becomes approximately conformal and $\Peq$ asymptotes to $\E/3$.
The bulk viscosity-to-entropy ratio as a function of temperature is shown in \fig{zeta}(top). In this case approximate conformal invariance implies that $\zeta/s \to 0$  at high and low  temperatures. In between,  $\zeta/s$ attains a maximum at 
\mbox{$T =0.22  \Lambda$}, reflecting the fact that the theory is maximally non-conformal around the scale set by the source.

\section{Collisions}
Remarkably, a gravitational wave propagating in the background \eqq{all} can be constructed simply by adding a term of the form 
\be
f(u) h(z_\pm) dz_\pm^2
\ee
to the metric \eqq{eq:dsvac}, with 
\bea
h(z_\pm) &=& \frac{\mu^3}{w \sqrt{2\pi}}  \, e^{-z_\pm^2/2\omega^2} \,, \\
f(u) &=& 4 \, e^{2A(u)} \,  \int_0^u \frac{d\tilde u}{\tilde u} \, e^{-4A(\tilde u)} \,.
\eea
The dual stress tensor of a single shock has
\be
\E= \Pl =h(z_\pm) \sac \Pt=0 \,. 
\ee
The scalar expectation value remains $\V=0$, as in the vacuum. The energy per unit transverse area of the shock is 
\be
\frac{d\E}{d^2 {\bf x}_\perp} = \mu^3 \,.
\ee
Unlike in the conformal case, in which the physics only depends on the dimensionless ``thickness'' $\mu \omega$ \cite{Casalderrey-Solana:2013aba}, in the present case the physics depends also on the initial transverse energy density in units of the source, $\mu/\Lambda$.  
We simulate collisions of $1/2\,$-shocks and $1/4\,$-shocks  in the terminology of \cite{Casalderrey-Solana:2013aba} ($\mu \omega=0.30$ and $\mu \omega=0.12$, respectively) for several different values of  $\mu/\Lambda$. We then extract the boundary stress tensor and we focus on its value at mid-rapidity, $z=0$, as a function of time.\footnote{We employ as a regulator a background thermal bath with an energy density between 0.8\% and 2.5\% of that at the centre of the initial shocks. We simulate each collisions with several different regulators and extrapolate to zero regulator.} 
We choose $t=0$ as the time at which the two shocks would have exactly overlapped in the absence of interactions \cite{Casalderrey-Solana:2013aba}.  

We define the hydrodynamization time, $\thy$, as the time beyond which both pressures are correctly predicted by the constitutive relations of first-order viscous hydrodynamics, 
\bse
\label{hydro}
\bal
\Pl^\textrm{hyd}  &= \Peq + \P_\eta + \P_\zeta \,, \\
\Pt^\textrm{hyd}  &= \Peq - \tfrac{1}{2}  \P_\eta + \P_\zeta \,,\end{align}
\ese
with a 10\% accuracy, so that  
\be
\frac{\left| {P}_{L,T} - {P}_{L,T}^\textrm{hyd} \right|}{\bar \P} <0.1\,.
\ee
In \eqq{hydro} we have denoted by $\P_\eta$ and $\P_\zeta$ the shear and the bulk contributions to the hydrodynamic pressures, respectively, which are proportional to the corresponding viscosities. The different coefficients in front of $\P_\eta$ in these two equations reflect the tracelessness of the shear tensor. We define the EoSization  time, 
$\teq$, as the time beyond which the average pressure coincides with the equilibrium pressure with a 10\% accuracy, meaning that  
\be
\frac{\left| \bar \P - \Peq \right|}{\bar \P} <0.1 \,. 
\ee

We expect on physical grounds that increasing the initial energy in the shocks increases the energy deposited in, and hence the hydrodynamization temperature of, the resulting plasma. We have confirmed that, indeed, $\Thy/\Lambda$ increases monotonically with $\mu/\Lambda$. On the gravity side this means that, for sufficiently large (small) $\mu/\Lambda$, the horizon forms in the UV (IR) region of the solution, where the geometry is approximately AdS. As a consequence, in these two limits the plasma formation and subsequent relaxation proceed approximately as in a CFT. In contrast, for $\mu \sim \Lambda$ the relaxation of the plasma takes place in the most non-conformal region where the bulk viscosity effects are largest. In this intermediate region we see several  effects that are absent in a CFT. 

First, hydrodynamization times are longer than in a CFT. This is illustrated by the dashed, red curve in \fig{zeta}(bottom) whose maximum, indicated by the first vertical line from the left, is 2.5 times larger than the conformal result, which is indicated  by the  horizontal  line.\footnote{This value differs from that in \cite{Casalderrey-Solana:2013aba} because 
\cite{Casalderrey-Solana:2013aba} used a 20\% criterion to define $\thy$. \label{note}}  
As expected, at high 
$\Thy/\Lambda$ we see that  $\thy \Thy$ asymptotically approaches its conformal value (we have checked that at $\Thy/\Lambda=4.8$ the difference is  0.5\%). We expect the same to be true at low $\Thy/\Lambda$.\footnote{Although we have not been able to verify this explicitly because simulations in this regime become increasingly challenging, \fig{zeta}(bottom) is  consistent with this expectation. \label{although}} 
The increase in the 
hydrodynamization time is qualitatively consistent with the increase in the lifetime of non-hydrodynamic quasi-normal modes found in  \cite{Buchel:2015saa,Janik:2015waa,Buchel:2015ofa,Rougemont:2015wca,Gursoy:2015nza,Gursoy:2016tgf,Attems:2016ugt,Janik:2016btb}. A heuristic explanation on the gravity side comes from realizing that the larger the non-conformality, the steeper the scalar potential becomes. As the plasma expands and cools down, the horizon ``rolls down the potential''. It is therefore intuitive that steeper potentials make it harder for the non-hydrodynamic perturbations of the horizon to decay. 

\begin{figure}[t]
\begin{center}
\includegraphics[width=.55\textwidth]
{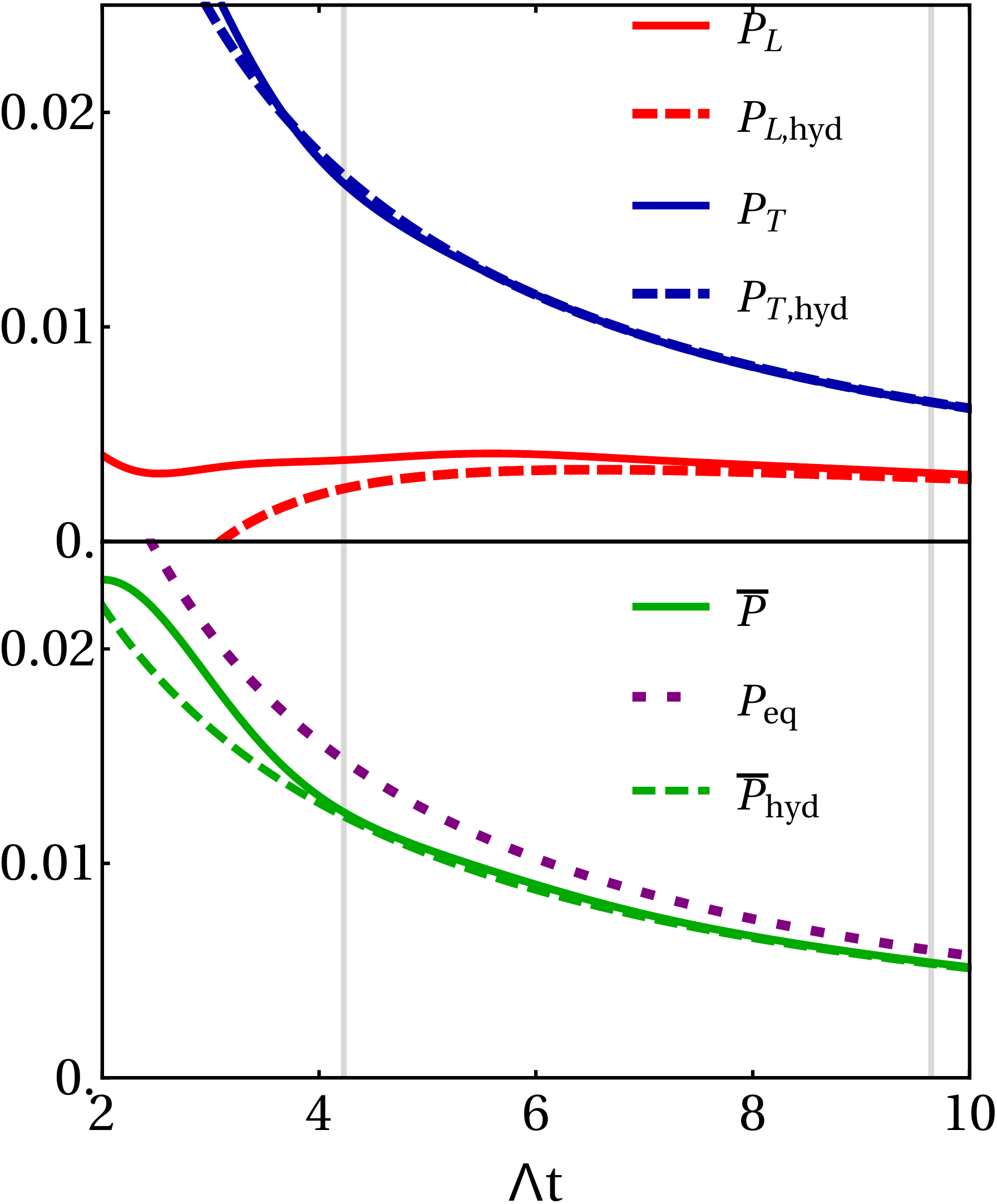}
\caption{\label{pressures}  Longitudinal, transverse and  average pressures, their  hydrodynamic approximations, and the equilibrium pressure extracted from the equation of state, all in units of $\Lambda^4$, for a collision of \mbox{$1/4\,$-shocks} with 
\mbox{$\mu / \Lambda = 0.94$}. The hydrodynamization temperature is $\Thy/\Lambda =0.24$. Because the transverse pressure hydrodynamizes  much faster than the longitudinal one, $\Pt$ and $\Pt^\textrm{hyd}$ are virtually on top of one another for the times shown. Hydrodynamization and EoSization take place at 
\mbox{$\thy \Lambda = 4.2$} and 
\mbox{$\teq \Lambda = 9.6$}, respectively, as indicated by the  vertical  lines. At $\thy$ the difference between $\bar \P$ and $\Peq$ is  18\%, whereas the difference between $\bar \P$ and 
${\bar \P}_\textrm{hyd}$ is  2\%. At $\teq$ the difference between $\Pl$ and $\Pl^\textrm{hyd}$ is 3\%. The $\Pt / \Pl$ ratio is 4.4 at $\thy$ and 1.9 at $\teq$.}
\end{center}
\end{figure}

Second, the equation of state is not obeyed out of equilibrium. This is illustrated 
in \fig{pressures}(bottom) for a collision of $1/4\,$-shocks with $\mu/\Lambda =0.94$, for which the  hydrodynamization temperature is $\Thy/\Lambda =0.24$. 
We see that the  equilibrium  and the  average pressures are not within 10\% of one another until a time 
\mbox{$\teq  = 9.6/\Lambda=2.4 / \Thy$}. This is further illustrated in \fig{zeta}(bottom), which  shows the dependence of the EoSization time on the hydrodynamization temperature for $1/2\,$-collisions. We see that for sufficiently large $\mu/\Lambda$ the EoSization time becomes negative, meaning that the  average and the equilibrium pressures  differ by less than 10\% even before the shocks collide. The reason is simply that in these cases the energy density in the Gaussian tails in front of the shocks, which start to overlap at negative times, becomes much higher than $\Lambda$. At these energy densities the physics becomes approximately conformal and  the equation of state becomes approximately valid as a consequence of this symmetry. An analogous argument implies that $\teq$ should also become negative for collisions with sufficiently small $\mu/\Lambda$.$^{\ref{although}}$

Third, hydrodynamization can take place before EoSization. Indeed, we see in \fig{zeta}(bottom) that \mbox{$\thy < \teq$} for collisions for which  the hydrodynamization temperature is between the first and the fourth vertical line.
Comparing with \fig{zeta}(top) we see that at these two temperatures the viscosity-to-entropy ratios are $\zeta/s = 0.025$ and $\zeta/s=0.017$, respectively. Note that the first value of $\zeta/s$ would decrease if we were to consider the lower temperature at which we expect that the two curves in \fig{zeta}(bottom) will have a second crossing. Also, note that the ordering of $\thy$ and $\teq$ depends on the  accumulated effect of the bulk viscosity along the entire history of the collision. Notwithstanding these caveats, we will take the value $\zeta/s = 0.025$ as a conservative estimate of the minimum bulk viscosity needed to have $\thy < \teq$ for $1/2\,$-collisions. The maximum value of the ratio $\teq/\thy$ for $1/2\,$-collisions  is $\teq/\thy = 2.56$. 

Regardless of the ordering of $\teq$ and $\thy$, these  times are always  shorter than the isotropization time beyond which $\Pl$ and $\Pt$ differ from one another by less than 10\%. This is apparent in  \fig{pressures}.

\section{Discussion}
 Eqs.~\eqq{hydro} imply that the hydrodynamic viscous correction to the equilibrium pressure  is controlled by the bulk viscosity alone, since 
\be
{\bar \P}_\textrm{hyd} = \Peq + \P_\zeta \,,
\label{bulk}
\ee
whereas the viscous deviation from isotropy is controlled by the shear viscosity alone, since
\be
\Pl^\textrm{hyd} - \Pt^\textrm{hyd} = \tfrac{3}{2} \P_\eta \,.
\label{shear}
\ee
We see from \eqq{bulk} that the reason why hydrodynamization can take place before EoSization is because hydrodynamics becomes applicable at a time when bulk-viscosity corrections are still sizeable. This is illustrated in \fig{pressures} by the fact that hydrodynamics provides an excellent prediction (within 2\%) for  ${\bar \P}$ at $\thy$, whereas at this time ${\bar \P}$ and $\Peq$ still differ  by 18\%. The above statement is the analog of the fact that hydrodynamization can take place before isotropization because hydrodynamics becomes applicable at a time when shear-viscosity corrections are still sizeable \cite{Chesler:2010bi}. In our model the bulk viscosity is rather small compared to the shear viscosity, since 
$\zeta/\eta = 4 \pi \zeta / s \simeq 0.35$ at the temperature at which $\zeta$ attains its maximum value. Presumably this is the reason why the difference  between $\Pl$ and $\Pt$ at $\thy$ in  \fig{pressures} is much larger than that between 
${\bar \P}$ and $\Peq$. 

Our results indicate that relaxation in non-conformal theories follows two qualitatively different paths depending on the bulk viscosity. If  
$\zeta/s \lesssim (\zeta/s)_\textrm{cross}$ then EoSization precedes hydrodynamization, whereas for 
\mbox{$\zeta/s \gtrsim (\zeta/s)_\textrm{cross}$} the order is reversed. Although we may take the cross-over value $(\zeta/s)_\textrm{cross} \sim 0.025$ obtained from $1/2\,$-collisions as representative, we emphasize that this depends not just on the model but on the specific  flow under consideration. For example, we expect 
$(\zeta/s)_\textrm{cross}$ to take a smaller value for $1/4\,$-collisions since in this case the gradients are larger than for  $1/2\,$-collisions \cite{Casalderrey-Solana:2013aba}. 
Note that along either of these paths, correlation functions, such as two point functions, may still differ from their thermal values, as explicitly demonstrated in
 \cite{Chesler:2012zk,Chesler:2011ds}.

Heavy ion collisions provide an excellent laboratory in which to study experimentally these two paths to relaxation. 
Indeed, although at very high temperatures the deconfined phase of QCD is approximately conformal, with very small values of $\zeta/s$, estimates of this ratio indicate that, in the vicinity of the critical temperature, $T_c$, a fast rise takes place to values as large as $\zeta/s \simeq 0.3$ \cite{Karsch:2007jc}. Despite the fact that  $\zeta/s$ is only sizeable  in a relatively narrow region around $T_c$, it has been shown to have an important effect on the late-time hydrodynamic description of the QGP created at RHIC and the LHC \cite{Torrieri:2008ip,Bozek:2011ua,Bozek:2012qs,Monnai:2009ad,Denicol:2010tr,Song:2009rh,Dusling:2011fd,Noronha-Hostler:2014dqa,Ryu:2015vwa,Florkowski:2015dmm}. Our results suggest  that the value of $\zeta/s$ may also have an impact on the early-time process of hydrodynamization. 
This may be investigated by comparing collisions of different systems  with  varying  energies. 
For most central collisions at top RHIC or LHC energies, the initial temperature is well above $T_c$ and hydrodynamization proceeds as in a conformal theory. However, in peripheral collisions or in central collisions at lower energies, such as those at the RHIC energy scan, the hydrodynamization temperature is reduced and the corresponding bulk viscosity may be sufficiently large to delay EoSization until after hydrodynamization. Another exciting possibility is to consider collisions of smaller  systems, such as $p-Pb$, $d-Au$, $^{3}He-Au$ or $p-p$ collisions. As it has been recently discovered, these systems also show strong collective behaviour \cite{Abelev:2012ola,Aad:2012gla,Adare:2013piz,Chatrchyan:2013nka,Adare:2015ctn,Adare:2015ctn,Aad:2015gqa}
 that is well described by hydrodynamic simulations \cite{Bozek:2011if,Nagle:2013lja,Schenke:2014zha,Kozlov:2014fqa,Romatschke:2015gxa,Bozek:2015qpa,Habich:2015rtj,Koop:2015trj} that include non-zero values of the bulk viscosity (see \cite{Dusling:2015gta} for a review of collective effects in this type of collisions). As stressed in \cite{Romatschke:2015gxa}, the temperature range explored by these smaller systems is narrowly concentrated around $T_c$. 
 These makes them ideal candidates with which to explore the effect of transport coefficients, in particular of the bulk viscosity.  
The comparison between the early-time dynamics of these small systems and heavy ion collision is an excellent framework in which to explore the different relaxation paths uncovered here. 

We close with a comment on on-going work. In order to ascertain the robustness of our conclusions, we are currently extending our simulations to other non-conformal models. In particular, we are  considering models that, unlike the one studied here, exhibit thermal phase transitions, near which non-hydrodynamics modes are expected to play an important role \cite{Janik:2015waa,Janik:2016btb}.
 
\section*{Acknowledgements}
We thank  P.~Figueras, N.~Gushterov, M.~Heller, 
I.~Papadimitriou,  U.~Sperhake,
A.~Starinets, W. van der Schee and L. G. Yaffe for discussions.
We thank the MareNostrum supercomputer at the BSC for computational  resources (project no.~UB65).
 MA is supported by the Marie Skodowska-Curie Individual Fellowship 
of the European CommissionÕs Horizon 2020 Programme under contract number 
658574 FastTh. JCS is a Royal Society University Research Fellow and was also supported by a Ram\'on y Cajal fellowship, by the Marie Curie Career Integration Grant FP7-PEOPLE-2012-GIG-333786 and by the Spanish MINECO through grant FPA2013-40360-ERC. CFS and DS acknowledge Spanish support from contracts AYA-2010-15709
(MICINN) and ESP2013-47637-P
(MINECO). DS acknowledges support via a FPI doctoral contract BES-2012-057909 from MINECO. We are also supported by grants MEC FPA2013-46570-C2-1-P, MEC FPA2013-46570-C2-2-P, MDM-2014-0369 of ICCUB, 2014-SGR-104, 2014-SGR-1474, CPAN CSD2007-00042 Consolider-Ingenio 2010, and ERC Starting Grant HoloLHC-306605.


\end{document}